\documentclass[num-refs]{wiley-article}
\usepackage{siunitx}

\papertype{Full paper}
\paperfield{Submitted to Magnetic Resonance in Medicine}

\title{Introduction to a novel T\textsubscript{2} relaxation analysis
method SAME-ECOS: Spectrum Analysis for Multiple Exponentials via
Experimental Condition Oriented Simulation}

\author[1,2\authfn{1}]{Hanwen Liu}
\author[1,3\authfn{0}]{Qing-San Xiang}
\author[3,4\authfn{0}]{Roger Tam}
\author[3]{Piotr Kozlowski}
\author[3\authfn{0}]{David K.B. Li}
\author[1,3\authfn{0}]{Alex L. MacKay}
\author[2,5\authfn{0}]{John K. Kramer}
\author[1,2,3,6\authfn{0}]{Cornelia Laule}

\affil[1]{Physics \& Astronomy, University of British Columbia}
\affil[2]{International Collaboration on Repair Discoveries}
\affil[3]{Radiology, University of British Columbia}
\affil[4]{Biomedical Engineering, University of British Columbia}
\affil[5]{Kinesiology, University of British Columbia}
\affil[6]{Pathology, University of British Columbia}

\corraddress{818 W 10th Ave, Vancouver, BC, V5Z1M9}
\corremail{hanwen@phas.ubc.ca}

\presentadd[\authfn{1}]{818 W 10th Ave, Vancouver, BC, V5Z1M9}

\fundinginfo{Funding support was provided by the Multiple Sclerosis Society of Canada, Natural Sciences and Engineering Research
Council Discovery Grant.}

\runningauthor{Liu et al.}

\begin{document}

\maketitle

\begin{abstract}

We propose a novel T\textsubscript{2} relaxation data analysis method whi-ch we have named spectrum analysis for multiple exponentials via experimental condition oriented simulation (SAME-ECOS).
%\textbf{S}pectrum \textbf{A}nalysis for \textbf{M}ulti-\textbf{E}xponentials via \textbf{E}xperimental \textbf{C}ondition \textbf{O}riented \textbf{S}imulation (\textbf{SAME-ECOS}).  
SAME-ECOS, which was developed based on a combination of information theory and machine learning neural network algorithms, is tailored for different MR experimental conditions, decomposing multi-exponential decay data into T\textsubscript{2} spectra, which had been considered an ill-posed problem using conventional fitting algorithms, including the commonly used non-negative least squares (NNLS) method. Our results demonstrated that, compared with NNLS, the simulation-derived SAME-ECOS model yields much more reliable T\textsubscript{2} spectra in a dramatically shorter time, increasing the feasibility of multi-component T\textsubscript{2} decay analysis in clinical settings.

% Please include a maximum of seven keywords
\keywords{SAME-ECOS, T\textsubscript{2} relaxation, non-negative least squares, resolution limit, myelin water imaging, machine learning}
\end{abstract}

\section{Introduction}
T\textsubscript{2} relaxation in biological tissues measured with a
multi-echo experiment is typically characterized by multi-exponenti-al
decays because multiple water pools may exist within a single image
voxel. \cite{whittall1999mono} The T\textsubscript{2} times of different
water pools are governed by the microenvironment of water molecules. For
example, myelin water, the water trapped in myelin bilayers, exhibits a
shorter T\textsubscript{2} relaxation time than that of
intra/extra-cellular water (IE) and free water. \cite{mackay1994vivo}
However, accurately depicting the spectrum of constituent
T\textsubscript{2} components for each image voxel from MR relaxation
data is nontrivial. Because the decay process of multiple water pools
takes place simultaneously, the MR receiver coil can only record a
signal that is the sum of multiple exponential decay components.
Consequently, one has to solve a mathematical problem of fitting a
superimposed relaxation signal into its constituent components. This
mathematically complex problem is commonly seen in many other
quantitative sciences and is often considered as an ill-posed
problem. \cite{istratov1999exponential} The analytical and numerical solutions to
this problem have been comprehensively reviewed by Istratov et al. from
a mathematical perspective. \cite{istratov1999exponential}

To provide a suitable solution specific to MR data, Whittall and
Mackay \cite{whittall1989quantitative} introduced the non-negative least squares
(NNLS) \cite{lawson1995solving,provencher1982contin,kroeker1986analysis} method that decomposes the multi-echo decay
data into a spectrum of positive T\textsubscript{2} times. The NNLS
method makes no prior assumptions about the total number of
T\textsubscript{2} components, which is a desirable feature for modeling
complex biological tissues with heterogeneous compositions. However,
without constraining the number of components, the T\textsubscript{2}
fitting problem becomes underdetermined with non-unique solutions,
making NNLS unstable and highly susceptible to noise even with strong
regularization. In contrast, other fitting methods such as the
quasi-Newton algorithm by Du et al. \cite{du2007fast} and the Wald
distribution by Akhondi-Asl et al. \cite{akhondi2014t} usually provide
more stable results and better noise resistance, but at the expense of
modelling only two or three water pools, which limits their usage on
unknown pathological tissues.

Simultaneously increasing the complexity and stability of a model is a
paradox, seemingly impossible to improve one without diminishing the
other in the game of multi-exponential decomposition. Our quest now was
to find the line of best balance between these two factors. According to
studies of information theory, the decay components can only be resolved
to a certain resolution limit at a given signal to noise ratio
(SNR). \cite{bertero1982recovery,mcwhirter1978numerical} That is to say, due to noise
contamination, there is always a limit in how closely the two
neighboring components can be resolved, regardless of how sophisticated
the analysis method is. This fundamental restriction on the resolution
limit leads to a correlation between the SNR and the maximum number of
components that can be reliably resolved in a particular analysis
range. \cite{link1991analysis} The exact expression of this correlation is
presented in the Methods, and plays a crucial role in our proposed
approach.

On the other hand, machine learning algorithms, in particular supervised
neural network methods \cite{schmidhuber2015deep}, have been successfully
implemented in many MR applications \cite{lundervold2019overview}, especially for
tasks involving parameter estimation. \cite{cohen2018mr,liu2020myelin_NN} In short, a
neural network can be trained to discover hidden patterns in data and to
learn the mapping between two vector spaces. A trained neural network
usually outperforms most conventional methods in terms of better
accuracy and faster speed, particularly if the mapping is highly
nonlinear. Additional background on the approximation properties of
neural networks can be found elsewhere. \cite{hornik1989multilayer}

\begin{figure}
    \centering
    \includegraphics[width=13cm]{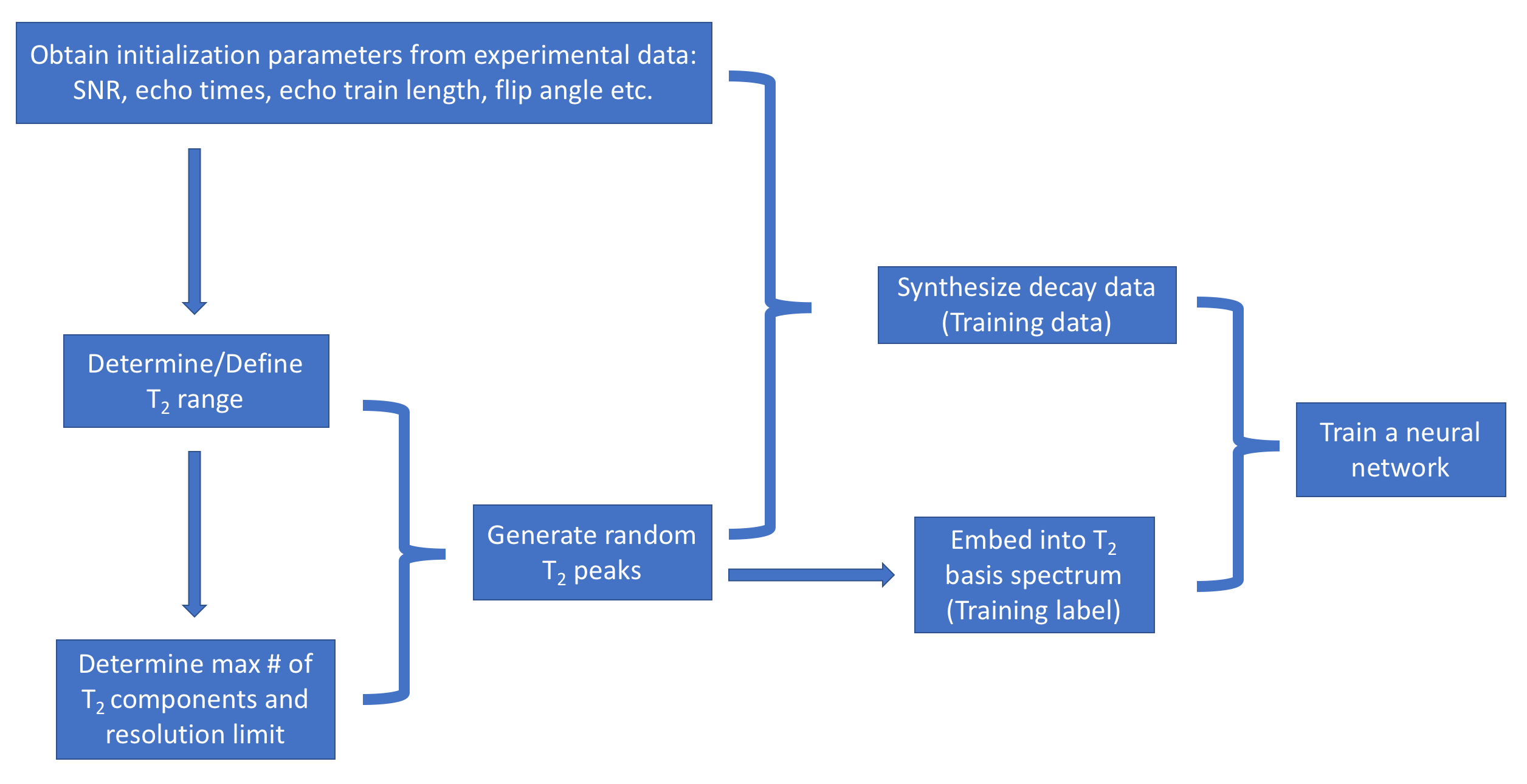}
    \caption{SAME-ECOS workflow.}
    \label{fig:SAME-ECOS workflow}
\end{figure}

Based on information theory and neural network algorithm approaches, we
propose a novel method which we have called \textbf{S}pectrum
\textbf{A}nalysis for \textbf{M}ulti-\textbf{E}xponentials via
\textbf{E}xperimental \textbf{C}ondition \textbf{O}riented
\textbf{S}imulation (\textbf{SAME-ECOS}) for the analysis of multi-echo
T\textsubscript{2} relaxation data. The general concept of SAME-ECOS (\textbf{Figure \ref{fig:SAME-ECOS workflow}})
{}
can be briefly described by a series of calculation,
simulation and model training operations:
(1) determine the T\textsubscript{2} range and resolution limit based on
the experimental conditions such as SNR and echo times;
(2) generate sufficient examples of random T\textsubscript{2} spectra
within the T\textsubscript{2} range obeying the resolution limit;
(3) compute multi-echo decay data using the randomly generated
T\textsubscript{2} spectra;
(4) train a neural network model to learn the mapping between the
simulated multi-echo decay data and the ground truth spectrum labels;
(5) apply the trained neural network model to experimental data.
In general, SAME-ECOS is a simulation-derived solver, powered by
information theory and machine learning, to the problem of fitting
multi-exponential decay data into a T\textsubscript{2} spectrum. It is
worth highlighting that SAME-ECOS has high flexibility of tailoring
itself to different experimental conditions attributed to its unique
simulation workflow. The detailed SAME-ECOS algorithm is demonstrated in
the Methods part by presenting an example with in-depth explanations.

\section{Methods}
Because the SAME-ECOS analysis method is tailored to different MR
experimental conditions, we use one specific example here as a paradigm
to demonstrate its simulation workflow, trained model evaluation, and
experimental data application. Therefore, the Methods part of this
introductory paper to SAME-ECOS is presented in the following sections
as a particular experiment.

\subsection{In-vivo MRI experiment}
32-echo brain data (gradient and spin echo (GRASE), TE/$\Delta$TE/TR=10/10/1000ms, refocusing flip angle (FA)=180$^{\circ}$, axial matrix size = 232$\times$186, acquired resolution = 1$\times$1$\times$5 mm\textsuperscript{3} for 20 slices, reconstructed resolution = 1$\times$1$\times$2.5 mm\textsuperscript{3} for 40 slices, acquisition time = 14.4 minutes) \cite{prasloski2012rapid} from one healthy volunteer (male, 57 years old) was collected at a 3T scanner (Philips Achieva) using an 8-channel head coil.

\subsection{SAME-ECOS workflow}
\label{section:SAME-ECOS workflow}
\textbf{(1) Choose the SNR range to be 70-300.} To initialize the
simulation workflow, the SNR of the experimental data needs to be
determined first, as calculations of the T\textsubscript{2} range, the
resolution limit, and especially the noise simulation, all depend on the
SNR. The SNR was initially estimated to be approximately 167 by
examining the noise variance of the air voxels proximate to the skull on
the 1\textsuperscript{st} echo image. However, most of today's MRI
images acquired by multi-channel coils are reconstructed using parallel
imaging \cite{deshmane2012parallel}, which complicates the estimation of SNR.
Other factors, such as B1 inhomogeneity, may lead to regional variations
in SNR. Therefore, the SNR of experimental data cannot be simply
assessed by a single definitive number. To accommodate these hurdles, we
empirically assigned the SNR a wide range of 70 to 300, instead of using
a single value. This approach allows us to randomly select any SNR
within the designated range at each simulation realization, making the
resulting simulated data `all-inclusive' after many realizations.

\textbf{(2) Define the T\textsubscript{2} range to be
7-2000ms.} Intuitively, the shortest detectable T\textsubscript{2} decay
component should have its residual signal greater than the noise level
at the first measurement. Because the first echo is used later as a
normalization factor, the second echo is actually regarded as the first
measurement in our analysis. In theory, a minimum of two measurement
points would be needed for the purpose of T\textsubscript{2} fitting.
That means the residual signal of the shortest T\textsubscript{2}
component in our analysis should be higher than the noise level at the
third echo. Thus, the lower bound of the T\textsubscript{2} range can
thus be obtained using equation \ref{eqn:lower boundary}.
\begin{equation}
\label{eqn:lower boundary}
T_{2}^{\min}=-\frac{3rd\ echo\ time}{\ln \left(\frac{1}{SNR}\right)}
\end{equation}
Given the SNR range of 70 to 300, the lower boundary is calculated to be
approximately 7ms. On the other hand, an ideal experimental condition
would be monitoring the decay as long as possible until the longest
T\textsubscript{2} component decays completely. \cite{shapiro1984use,dobaczewski1994laplace,thomasson1974analysis}
Then the upper bound for the analyzable T\textsubscript{2} range can be
determined using equation \ref{eqn:upper boundary}. \cite{istratov1999exponential}
\begin{equation}
\label{eqn:upper boundary}
T_{2}^{\max}=-\frac{last\ echo\ time}{\ln \left(\frac{1}{SNR}\right)}
\end{equation}
However, the ideal experimental condition rarely happens as the decay
monitoring time is often compromised to achieve a shorter scanning time
for most in-vivo MR experiments including our own (last TE = 320ms). If
equation \ref{eqn:upper boundary} were used with our SNR range of 70 to 300, the longest
analyzable T\textsubscript{2} component would be less than 80ms, which
is considered to be too short for the analysis of in-vivo brain imaging.
Therefore, the upper boundary is manually extended to 2000ms based on
the literature T\textsubscript{2} ranges for
brain. \cite{whittall1999mono,mackay1994vivo,kroeker1986analysis}

\textbf{(3). Determine the maximum number of resolvable
T\textsubscript{2} components to be M = 5 and the resolution
limit $\delta$ = 3.098.} For a given SNR and T\textsubscript{2} range, the
decay components can only be resolved to a certain resolution limit.
Link et al. \cite{link1991analysis} derived an expression (equation \ref{eqn:SNR T2 range correlation})
relating \(T_{2}^{\min}\), \(T_{2}^{\max}\) and SNR, to the M
resolvable exponentials. 
\begin{equation}
\label{eqn:SNR T2 range correlation}
\frac{M}{\ln \left(\frac{T_{2}^{\max }}{T_{2}^{\min }}\right)} \times \sinh \left(\frac{\pi^{2} \times M}{\ln \left(\frac{T_{2}^{\max }}{T_{2}^{\min }}\right)}\right)=\left(\frac{S N R}{M}\right)^{2}
\end{equation}
 Derivations and justifications of equation \ref{eqn:SNR T2 range correlation}
can be found in several publications. \cite{istratov1999exponential,mcwhirter1978numerical,link1991analysis} 
Based on
these previous theoretical studies, the maximum number of resolvable
T\textsubscript{2} components M $=$ 5 was obtained by numerically
solving equation \ref{eqn:SNR T2 range correlation} to the nearest integer for the given SNR and
T\textsubscript{2} range. Note that M $=$ 5 is the universal integer
solution to most SNRs in the range of 70-300 in our case.
Then, the T\textsubscript{2} resolution limit $\delta$=3.098 was determined by
equation \ref{eqn:resolution limit}. \cite{istratov1999exponential}
\begin{equation}
\label{eqn:resolution limit}
\delta=\left(\frac{T_{2}^{\max }}{T_{2}^{\min }}\right)^{\frac{1}{M}}
\end{equation}

\textbf{(4). For each simulation realization:}

\textbf{a. generate random integers n $<$ M, FA $\in$ {[}90\textsuperscript{$\circ$}, 180\textsuperscript{$\circ$}{]}, and SNR $\in$ {[}70,300{]}.} Note that the simulation did not include the case of n $=$ M because when n $=$ M, there is only one possible configuration for the M T\textsubscript{2} locations within the T\textsubscript{2} range obeying
the resolution limit, which would introduce an unwanted bias into the
simulated dataset. Also note that the actual FA can deviate substantially from the
prescribed FA due to B1 inhomogeneity, so we gave a
90\textsuperscript{$\circ$} tolerance to account for the FA variations.

\textbf{b. generate n randomly located T\textsubscript{2} components
	within the T\textsubscript{2} range obeying the resolution limit $\delta$.} The locations and amplitudes of the n T\textsubscript{2} components were randomly assigned and normalized to one.

\textbf{c. synthesize 32-echo decay data.} The pure decay signal without
noise of the n T\textsubscript{2} components (denoted as
\(S_{\text{pure}}\)) were synthesized (equation \ref{eqn:EPG yield pure decay data}) using the extended
phase graph (EPG) algorithm \cite{hennig1988multiecho,prasloski2012applications} with T\textsubscript{1}
= 2000ms as a default.
\begin{equation}
\label{eqn:EPG yield pure decay data}
S_{\text {pure}}=\sum_{i=1}^{i=n} \text {Amplitude}_{i} \times EPG\left(T_{2, i}, \text { selected FA}\right)
\end{equation}
To mimic the noise profile of a real MRI image \cite{cardenas2008noise}, the \(S_{\text{pure}}\)
was first projected into the real and imaginary axes by a random phase
factor \(\theta\) $\in$ {[}0\textsuperscript{$\circ$}, 90\textsuperscript{$\circ$}{]},
followed by adding noises on both axes, and finally producing the
magnitude of the noisy signal (equation \ref{eqn: rotate signal and add noise}).
\begin{equation}
\label{eqn: rotate signal and add noise}
S_{\text{noisy}}=\sqrt{\left(S_{\text{pure}} \times \sin \theta+noise_{-} 1\right)^{2}+\left(S_{\text{pure}} \times \cos \theta+ noise_{-} 2\right)^{2}}
\end{equation}
where \(noise\_ 1\) and \(noise\_ 2\) were independently sampled from a
Gaussian distribution with its mean $=$ 0 and its variance was determined
by equation \ref{eqn: Gaussian noise variance}
\begin{equation}
\label{eqn: Gaussian noise variance}
\text {Gaussian noise variance}=\frac{1}{\text {selected SNR} \times \sqrt{\pi / 2}}
\end{equation}
such that the noisy signal \(S_{\text{noisy}}\) would follow a Rician
distribution at the selected SNR level. The synthesized noisy decay data
were subsequently normalized to the 1\textsuperscript{st} echo and saved
for model training.

\textbf{d. Embed n T\textsubscript{2} components into a spectrum
	representation depicted by 40 basis T\textsubscript{2}s}. The 40 basis
T\textsubscript{2}s (t\textsubscript{1}, t\textsubscript{2}
t\textsubscript{3} \ldots{} t\textsubscript{40}) were equally spaced
within the T\textsubscript{2} range on a logarithmic scale. The 40
weighting factors of the basis T\textsubscript{2}s were used to
represent the spectrum of n T\textsubscript{2} components
(T\textsubscript{2,1}, T\textsubscript{2,2}, \ldots{}
T\textsubscript{2,n}). Explicitly, each T\textsubscript{2} component was
depicted as a Gaussian-shaped peak by the basis T\textsubscript{2}s,
with the weighting factor \(w_{i}\ \)of the i\textsuperscript{th} basis
T\textsubscript{2} (t\textsubscript{i}) being calculated as
\begin{equation}
w_{i}=\sum_{j=1}^{n} \frac{1}{\sqrt{2 \pi}} e^{-\frac{1}{2}\left(t_{i}-T_{2, j}\right)^{2}}
\end{equation}
The embedded T2 basis representation was normalized to one and recorded as the ground truth spectra for model training.

\textbf{(5) Train a neural network to map the decay data to its
T\textsubscript{2} spectrum.} A neural network (hidden layers:
\(100 \times 500 \times 1000 \times 1000 \times 500\), activation:
SeLU \cite{klambauer2017self} (hidden layers) and softmax \cite{bishop2006pattern}
(output layer), optimizer: Adamax \cite{kingma2014adam}, loss: categorical
cross-entropy \cite{lecun2015deep}) was constructed using
TensorFlow \cite{abadi2016tensorflow} to take 32-echo decay data as input and
predict the weighting factors of the 40 basis T\textsubscript{2}s at the
output layer. The constructed neural network was trained to map the
decay data to its T\textsubscript{2} spectrum. We yielded 3,000,000
simulation realizations (step 4), 90\% of which were used for the neural
network training. The remaining 10\% simulation realizations were used
for the validation that determined the stopping criterion for the
training process. The training was stopped when the accuracy on the
validation set did not improve further. This particular trained neural
network is denoted as the SAME-ECOS model hereafter and can be applied
to new 32-echo decay data to obtain T\textsubscript{2} spectrum.

\subsection{SAME-ECOS model performance evaluation}

The performance of the SAME-ECOS model was evaluated using three
designed tests and compared respectively with the results determined by
a regularized NNLS solver equipped with stimulated echo
correction \cite{prasloski2012applications} (analysis program can be requested here:
\url{https://mriresearch.med.ubc.ca/news-projects/myelin-water-fraction/}).
The kernel matrix for the NNLS analysis was adjusted accordingly to
match our experimental and simulation parameters. The regularization
parameter was chosen to be the largest value that allows a misfit of
less than 1.02 times of the minimum misfit, which is commonly used in
many studies. \cite{dvorak2019rapid,liu2019myelin,liu2020myelin,lee2018inter,ljungberg2017rapid}

\textbf{Test 1:} 300,000 ground truth T\textsubscript{2} spectra and
their noisy 32-echo decay data were randomly generated following the
workflow described in section \ref{section:SAME-ECOS workflow}. The decay data were analyzed by the
SAME-ECOS model and NNLS respectively to produce the T\textsubscript{2}
spectra. The processing time was recorded. The `goodness' of each
spectrum fitting by both methods was quantitatively assessed using
cosine similarity scores \cite{han2011data}, which report values
between 0 to 1, with 0 being the least similar and 1 being the most
similar to the ground truth labels. The cosine similarity score is a
commonly used metric that measures the similarities between two vectors,
especially when the vectors are high dimensional. It is a suitable
metric for our task since each T\textsubscript{2} spectrum can be
treated as a vector of 40 dimensions. The calculation of the cosine
similarity score is defined explicitly in the following formula
\begin{equation}
cosine\ similarity\ score = \frac{X \bullet Y}{\|X\| \times\|Y\|}
\end{equation}
Where \(X\) and \(Y\) are the vector representations of the predicted
and the ground truth spectra; \(\parallel X \parallel\) and
\(\parallel Y \parallel\) are their Euclidean norms respectively. Paired
t-test was performed to determine whether there was a significant
difference (P\textless{}0.05) in the cosine similarity scores calculated
by SAME-ECOS and NNLS.

\textbf{Test 2:} To examine the model robustness to noise, simulated
decay data of 4 pre-defined ground truth T\textsubscript{2} spectra
(\textbf{Table \ref{table:pre-defined T2 spectra}}) at SNR = 100 and FA = 180\textsuperscript{$\circ$}, each with 100
different noise realizations, were passed to the SAME-ECOS model and
NNLS for spectrum predictions. 
\begin{table}
\centering
\begin{tabular}{|c|c|c|}
\hline
\textbf{Pre-defined spectra} & \textbf{T\textsubscript{2} locations (ms)} & \textbf{T\textsubscript{2} amplitudes (normalized)} \\ \hline
Spectrum 1                   & 100                         & 1                                   \\ \hline
Spectrum 2                   & 25, 120                    & 0.3, 0.7                            \\ \hline
Spectrum 3                   & 15, 80, 50                & 0.3, 0.5, 0.2                     \\ \hline
Spectrum 4                   & 10, 60, 300, 1200           & 0.2, 0.4, 0.3, 0.1                \\ \hline
\end{tabular}
\caption{\textbf{Four pre-defined ground truth spectra.} The T\textsubscript{2} locations of each spectrum are selected obeying the resolution limit. Amplitudes are normalized to one.}
\label{table:pre-defined T2 spectra}
\end{table}
The location and amplitude of
each ground truth spectrum were manually chosen to provide a
visual-friendly data presentation. The spectrum analysis results of both
SAME-ECOS and NNLS were normalized (sum to unity) prior to comparison.
The similarity between each predicted and ground truth spectrum was
assessed by cosine similarity scores defined above. The mean and
standard deviation of the cosine similarity scores were also calculated.

\textbf{Test 3:} 10,000 ground truth T\textsubscript{2} spectra and
their 32-echo decay data with noise realizations were randomly generated
according to steps described in section \ref{section:SAME-ECOS workflow}. The decay data were
analyzed by the SAME-ECOS model and NNLS respectively to produce the
T\textsubscript{2} spectra. The myelin water fraction (MWF, a fraction
of signal with T\textsubscript{2}s \textless{} 40ms)\textsuperscript{2}
was extracted from each predicted spectrum and compared with the ground
truth MWF. Mean absolute error (MAE) in the MWF estimation was computed.
The correlation between the errors of MWF estimation and the FA was
evaluated using Pearson correlation analysis.

\subsection{Apply SAME-ECOS to experimental data}
The SAME-ECOS model was applied to the experimental in-vivo GRASE data,
which were pre-processed by normalizing to the first echo image. The
processing time for the whole brain data analysis was recorded. From the
resulting T\textsubscript{2} spectra, the MWF (T\textsubscript{2}s
\textless{} 40ms) was extracted for each voxel. The masks of regions of
interest (ROI) including whole brain, whole white matter, corpus
callosum, corticospinal tract, forceps major, and forceps minor were
produced using the first echo image via the FSL segmentation
tool. \cite{zhang2001segmentation} The T\textsubscript{2} spectra and MWF map was
also produced using NNLS in the T\textsubscript{2} range of 7-2000ms as
a reference for comparison.

\section{Results}
\subsection{Performance evaluation via simulation tests}
\subsubsection{Test 1: general performance}
\label{section:SAME-ECOS test 1}
The processing time to make 300,000 spectra predictions was 22 seconds
for the SAME-ECOS model and 1,620 seconds for NNLS (CPU: Intel(R)
Core(TM) i7-5930K @ 3.5 GHz, 32 GB RAM). The mean cosine similarity
score of the 300,000 spectra predicted by the SAME-ECOS model
(0.838$\pm$0.189) was significantly higher (p\textless{}0.05) than that of the NNLS (0.741$\pm$0.160).

\subsubsection{Test 2: robustness to noise}
\label{section:SAME-ECOS test 2}
\begin{figure}
    \centering
    \includegraphics[height=17cm]{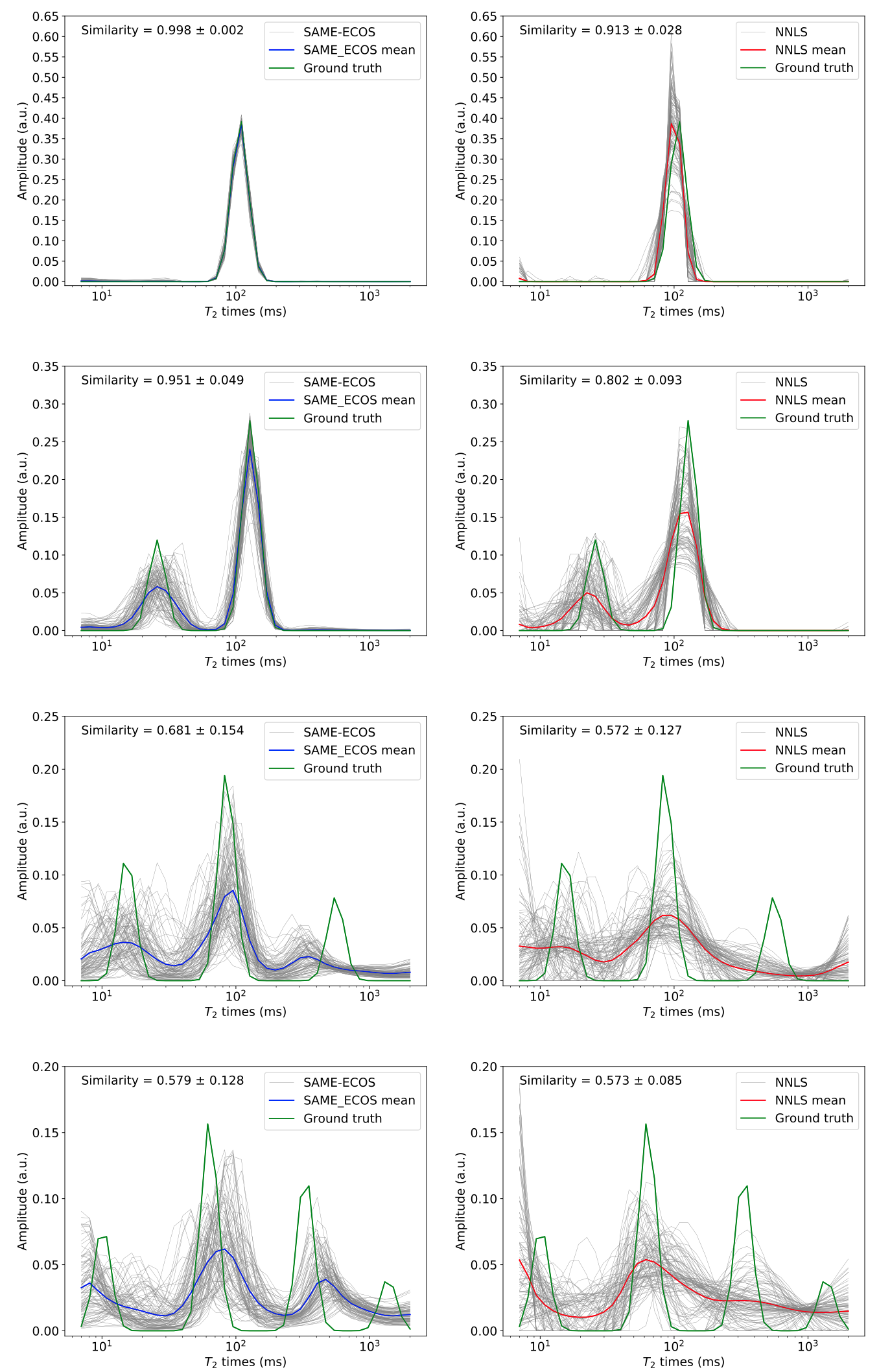}
    \caption{\textbf{T\textsubscript{2} spectra produced by SAME-ECOS (blue) and NNLS (red) respectively are compared with the ground truth spectra (green).} Simulated decay data, which are generated from 4 pre-defined ground truth spectra (\textbf{Table 1}) each with 100 different noise realizations, are fed into the trained SAME-ECOS model and NNLS algorithm to generate the T\textsubscript{2} spectra. Faded gray lines indicate the produced spectra for each noise realization. The mean and standard deviation of cosine similarity scores of 100 realizations are shown for each sub-figure.}
    \label{fig:spectra prediction comparison}
\end{figure}

The resulting SAME-ECOS and NNLS spectra from each individual noise
realization (grey) were plotted in \textbf{Figure \ref{fig:spectra prediction comparison}} to compare against
the ground truth spectrum (green). 
The average spectrum from 100
different noise realizations was also calculated for all SAME-ECOS and
NNLS scenarios (blue: SAME-ECOS, red: NNLS). SAME-ECOS produced visually
better results than NNLS for all scenarios. For spectra consisting of
one or two T\textsubscript{2} components, the SAME-ECOS model was able
to make almost perfect predictions (cosine similarity score: 0.998$\pm$0.002
and 0.951$\pm$0.05649, respectively); NNLS could also make accurate
predictions but with more substantial variability (cosine similarity
score: 0.913$\pm$0.0.028 and 0.802$\pm$0.093, respectively). For spectra
consisting of three or four T\textsubscript{2} components, the
performances of both methods started to degrade. However, SAME-ECOS
(cosine similarity score: 0.681$\pm$0.154, 0.579$\pm$0.128, respectively) was
still making better predictions than NNLS (cosine similarity score:
0.572$\pm$0.127, 0.573$\pm$0.085, respectively).

\subsubsection{Test 3: MWF prediction accuracy}
\label{section:SAME-ECOS test 3}
\begin{figure}
    \centering
    \includegraphics[width=10cm]{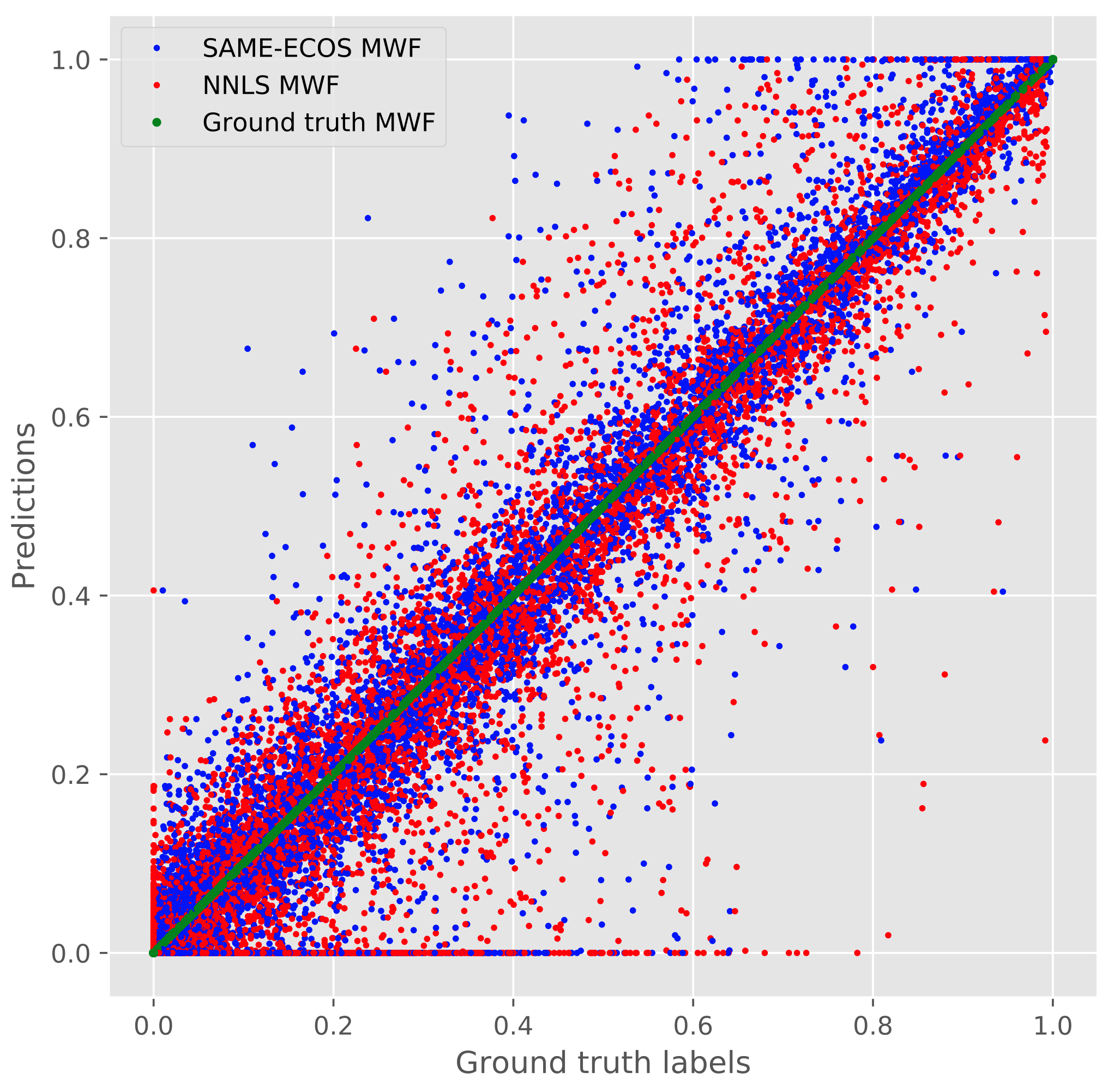}
    \caption{\textbf{Myelin water fraction (MWF) produced by SAME-ECOS (blue) and NNLS (red) are compared with the ground truth MWF (green).} 10,000 randomly simulated decay data examples were analyzed by SAME-ECOS and NNLS respectively. The MWF (fraction of signal with T\textsubscript{2}s \textless{} 40ms) was extracted from each predicted spectrum.}
    \label{fig:MWF SAME-ECOS vs. NNLS}
\end{figure}

\begin{figure}
    \centering
    \includegraphics[width=14cm]{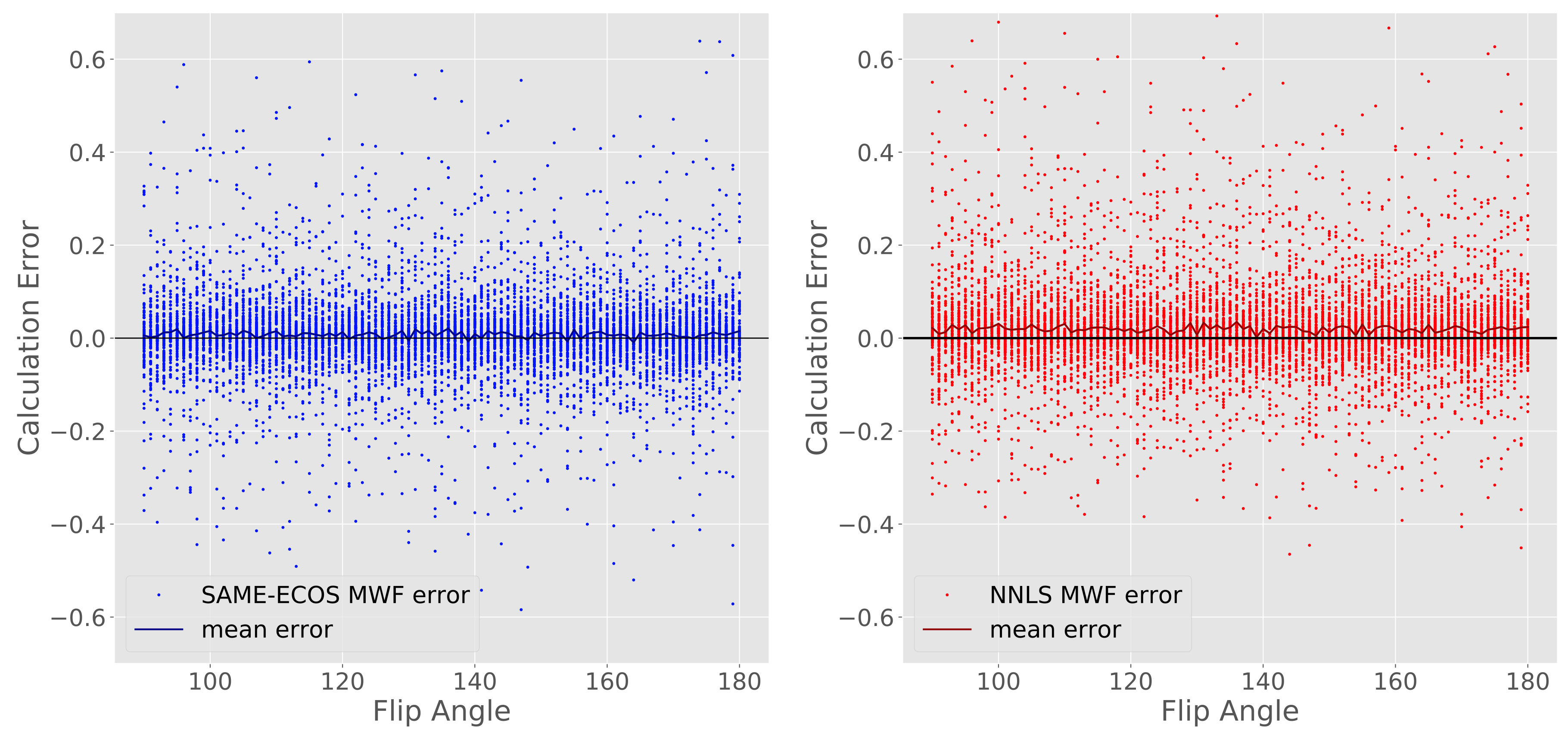}
    \caption{\textbf{Flip angle (FA) dependence of myelin water fraction (MWF) prediction errors.} 10,000 randomly simulated decay data examples were analyzed by SAME-ECOS and NNLS. Errors of MWF (fraction of signal with T\textsubscript{2}s \textless{} 40ms) was calculated and compared with FA. The mean error of each FA was also presented for a visual check for biases.}
    \label{fig:FA dependence}
\end{figure}

MWF values were extracted using both SAME-ECOS and NNLS methods and
plotted against 10,000 ground truth MWF values in
\textbf{Figure \ref{fig:MWF SAME-ECOS vs. NNLS}}. 
SAME-ECOS MWF (blue, MAE=0.050) demonstrated slightly better agreement
with the ground truth (green) than NNLS MWF (red, MAE=0.058). The MWF
prediction errors of both methods are plotted against the FA in \textbf{Figure \ref{fig:FA dependence}}, where the mean error of each FA was also presented
for a visual check for biases. A small but noticeable positive bias
(0.019, overestimation of MWF) was observed for the NNLS method, whereas
SAME-ECOS did not show any obvious bias (0.007). 

\subsection{In-vivo experimental data: MWF maps}
\begin{figure}
    \centering
    \includegraphics[width=14cm]{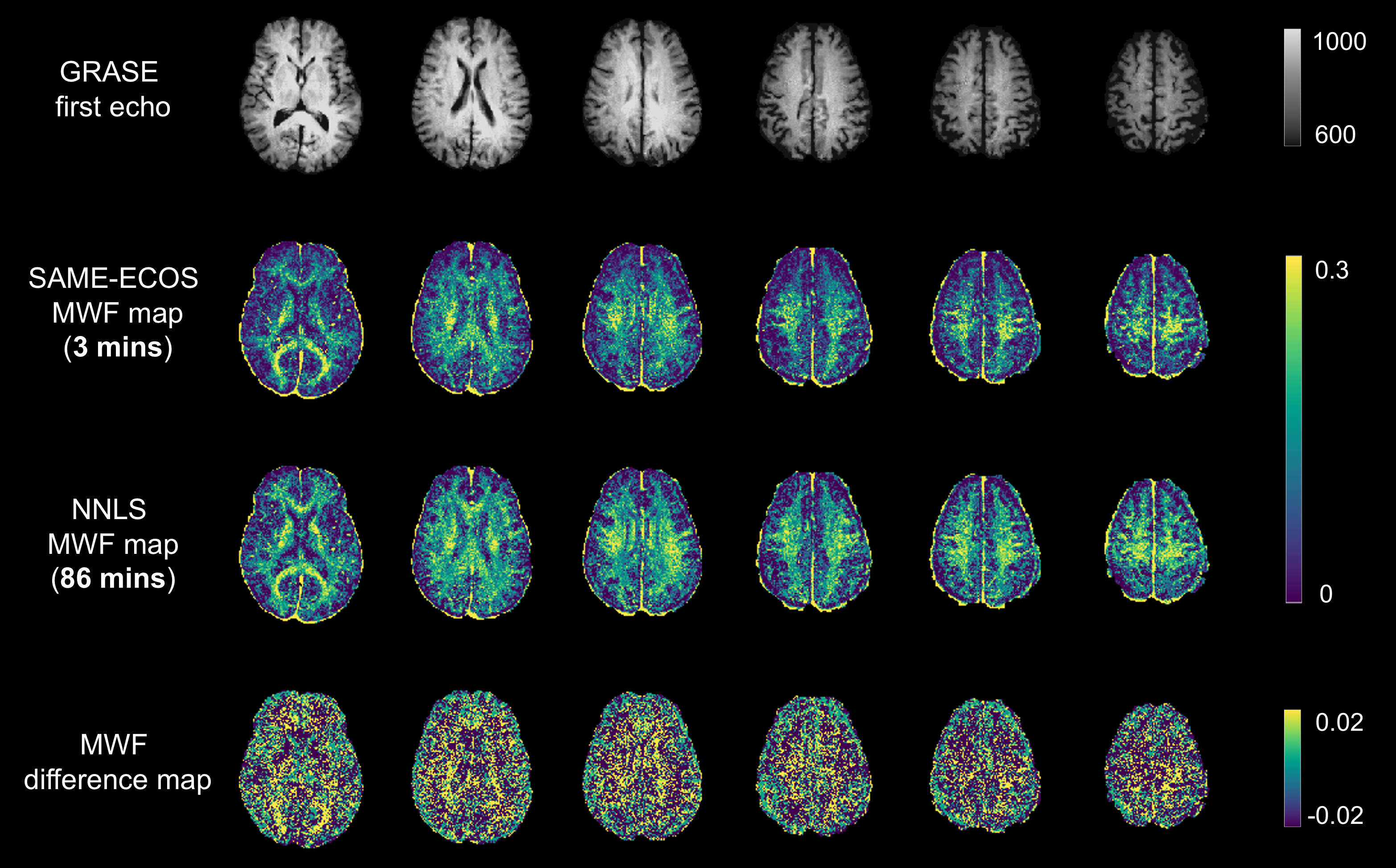}
    \caption{\textbf{SAME-ECOS and NNLS derived in-vivo myelin water fraction maps.} Six representative slices of the GRASE first echo, the resulting MWF maps produced by both methods (T\textsubscript{2}s \textless{} 40ms), and the voxel-wise MWF difference map (SAME-ECOS MWF \(-\) NNLS MWF), are shown.}
    \label{fig:in-vivo MWF maps}
\end{figure}

\begin{figure}
    \centering
    \includegraphics[height=17cm]{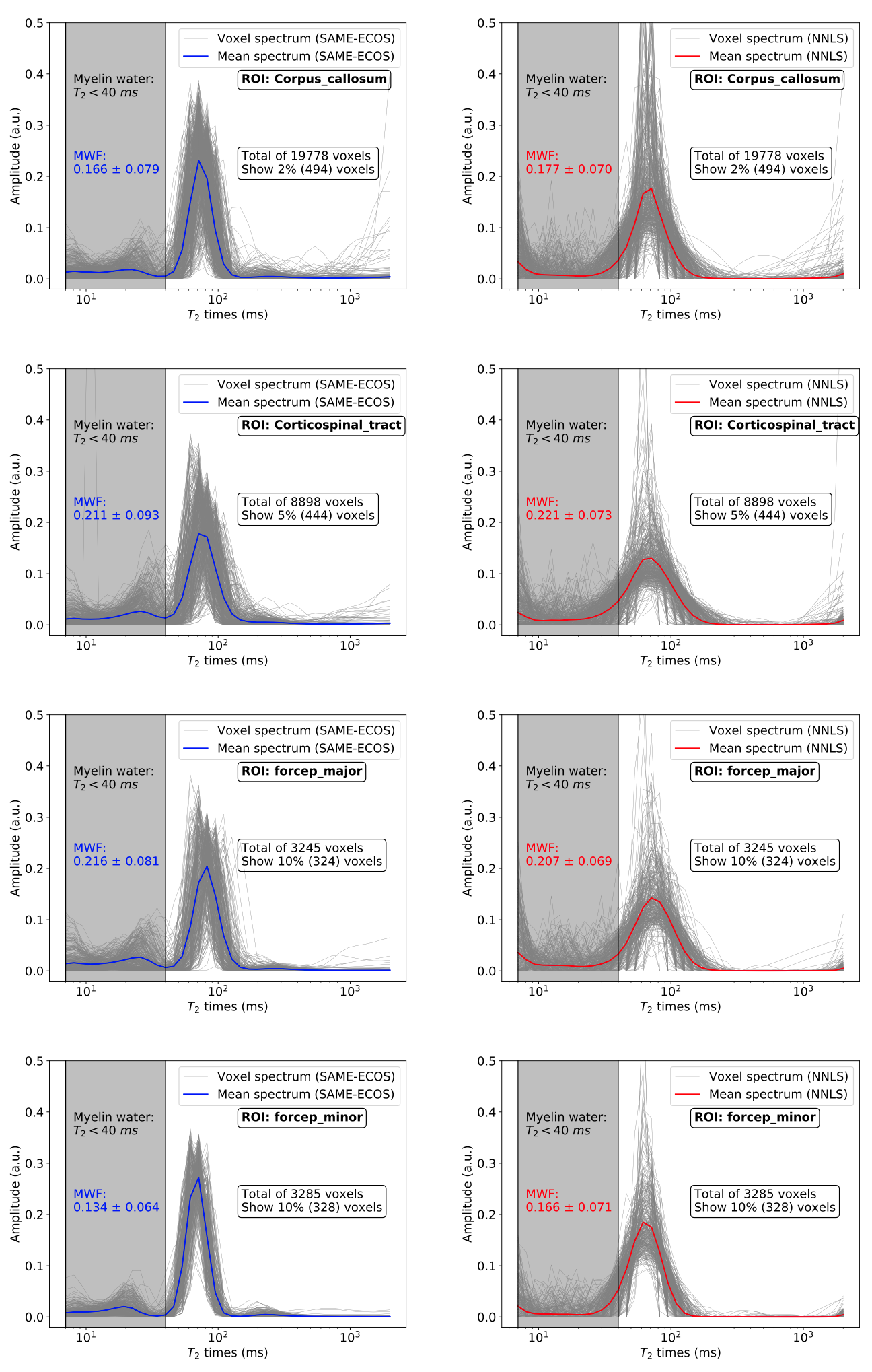}
    \caption{\textbf{Voxel spectra (gray), the mean spectra (blue: SAME-ECOS; red: NNLS), and the mean MWF (T\textsubscript{2}s \textless{} 40ms) within the four ROIs of the experimental in-vivo data.} ROIs include the corpus callosum, corticospinal tract, forceps major, and forceps minor. The voxel spectra are only plotted for a fraction of the total number of voxels for a visual-friendly presentation as indicated in each subfigure. The mean spectrum and mean MWF are calculated from all voxels within each ROI.}
    \label{fig:in-vivo spectrum}
\end{figure}

In-vivo GRASE data were analyzed by the SAME-ECOS model and NNLS. The
processing times of the whole brain data were 3 minutes for the
SAME-ECOS model and 86 minutes for NNLS. Six representative slices of
the GRASE first echo, the resulting MWF maps produced by both methods
(T\textsubscript{2}s \textless{} 40ms), and the voxel-wise MWF
difference map (SAME-ECOS MWF \(\  - \ \ \)NNLS MWF), are presented in \textbf{Figure \ref{fig:in-vivo MWF maps}}. 
The SAME-ECOS MWF map is visually very similar to the
NNLS MWF map, but subtle differences are still visible. Quantitatively,
the SAME-ECOS mean MWF of whole white matter (0.131$\pm$0.080) is lower than
that of the NNLS approach (0.152$\pm$0.078). \textbf{Figure \ref{fig:in-vivo spectrum}} shows the
voxel spectra, the mean spectra, and the mean MWF within the ROIs of the
corpus callosum, corticospinal tract, forceps major and forceps minor.
The SAME-ECOS mean MWF was lower than the NNLS mean MWF for most ROIs
(except for forceps major). It is observed that for all ROIs, the
SAME-ECOS mean spectra are able to resolve a short T\textsubscript{2}
peak (attributed to myelin water) in addition to a more dominate middle
peak (attributed to IE water), whereas the NNLS mean spectra can only
resolve a single broad middle peak.

\section{Discussion}
\subsection{SAME-ECOS vs. NNLS}
From the results of all three simulation tests, the SAME-ECOS model
largely outperformed NNLS. Using NNLS as the baseline, the SAME-ECOS
model achieved 13.1\% higher overall cosine similarity scores
(\textbf{Test 1}), and 16.0\% lower MAE of MWF predictions (\textbf{Test
3}), as well as demonstrated better robustness to noise (\textbf{Test
2}). Specifically, from visual inspection of the results presented in
\textbf{Figure 2}, both methods were able to produce accurate
T\textsubscript{2} spectra when the number of T\textsubscript{2}
components n $\leq$ 2. However, when n \textgreater{} 2, the NNLS predictions
became extremely unstable, resulting in over-smoothed mean spectra This
observation illustrates that, unlike NNLS being highly susceptible to
noise, the SAME-ECOS model has a desirable feature of being relatively
noise-inert. It is also noticed that both methods were incapable of
producing reliable spectrum predictions particularly for the long
T\textsubscript{2} components (e.g. T\textsubscript{2}s \textgreater{}
500ms), which is likely due to the relatively short measurement time
(the last TE=320ms) so there was not enough information acquired to
resolve the long T\textsubscript{2} component accurately. In addition,
as illustrated in \textbf{Figure 4}, NNLS was systematically
overestimating MWF throughout the FA range under our investigation. In
contrast, SAME-ECOS produced more accurate MWF values without any
obvious biases. Overall, the current SAME-ECOS model demonstrated a
better performance than NNLS in the interrogations using simulation
tests.

Due to the lack of ground truth, it is difficult to judge which method
is more accurate when it comes to in-vivo experimental data. However, in
terms of data processing speed, SAME-ECOS is approximately 30 times
faster than NNLS, achieving a whole-brain analysis in 3 minutes, which
is more feasible in clinical settings. From a visual inspection of the
MWF maps shown in \textbf{Figure 5}, these two methods produced similar
results, but the SAME-ECOS MWF map appeared to be a little less noisy,
making its white and gray matter distinction slightly more prominent.
Interestingly, compared with SAME-ECOS, the NNLS mean MWF values in
white matter regions were higher, which coincided with the NNLS MWF
overestimation issue presented in \textbf{Figure 4}.

A thorough analysis of the NNLS spectra in a few white matter ROIs
(\textbf{Figure 6}) revealed that the NNLS MWF signal
(T\textsubscript{2} cutoff at 40ms) largely originated from the broad
middle peak, which is commonly designated as the IE water pool. There
was not much contribution from a separate myelin water pool
(T\textsubscript{2}s \textless{} 40ms) to the NNLS MWF, and the myelin
water peak did not exist on the NNLS mean spectra for all ROIs. In
contrast, the myelin water peak and its separation with the IE water
peak were easily seen on the SAME-ECOS mean spectra for all ROIs
(\textbf{Figure 6}), aligning with the intuitive interpretation of MWF
arising entirely from a separate myelin water pool. Thus, this
observation raises a concern that reporting MWF values alone may not
fully unveil the underlying information, especially for those myelin
water imaging studies that used NNLS. Nevertheless, future studies using
histological validation \cite{laule2008myelin,laule2006myelin} should be conducted to
better compare these two analysis methods in dealing with experimental
data.

\subsection{Advantages of SAME-ECOS}

The biggest advantage of SAME-ECOS lies in the fact that it is a
simulation-derived solver, which takes a fundamentally different route
to yield the solutions compared with conventional solvers such as NNLS.
Its unique workflow makes SAME-ECOS highly tailorable to different
experimental conditions and the needs of analysis. Most parameters used
in SAME-ECOS are tunable. For example, the T\textsubscript{1} parameter
was set to be a constant for all T\textsubscript{2} components in our
simulation because the T\textsubscript{1} weighting is minimized by
using GRASE. But if a sequence sensitive to both T\textsubscript{1} and
T\textsubscript{2} were used, one could simply turn the
T\textsubscript{1} parameter into another variable for the production of
training data and make the training labels a 2D map that resolves
distinct components with different T\textsubscript{1},
T\textsubscript{2} times. \cite{celik2013stabilization} In contrast, it is
presumably more difficult to include either T\textsubscript{1} or any
other quantities as an additional variable in the NNLS analysis. Another
example would be how easily the influence of B1 inhomogeneity is
handled. SAME-ECOS simply treats the refocusing FA as a variable when
producing the simulated decay data to account for the FA variations
caused by B1 inhomogeneity. However, the same problem was not solved
using NNLS for many years until Prasloski et al. \cite{prasloski2012applications}
proposed to integrate the stimulated echo correction into the original
NNLS algorithm, which was a tremendous effort for such integration.
Furthermore, if one wants to subject the data analysis to a simpler
3-pool model for instance, SAME-ECOS can easily be simplified to a
3-pool solver by fixing the number of T\textsubscript{2} components to
be n=3 in the simulation, although it is not encouraged to do so for
reasons discussed later. Conclusively, SAME-ECOS is extremely tunable to
accommodate either a simpler or more complex model.

Another advantage of SAME-ECOS is to utilize the concept of resolution
limit from information theory to prevent overfitting of noise. Due to
unavoidable noise contamination, T\textsubscript{2} components can only
be recovered to a certain extent, and any violation to the law of
resolution limit should be theoretically prohibited. The essential part
of SAME-ECOS workflow is to obey the resolution limit to prevent any
solutions with adjacent T\textsubscript{2} components being too close
together, a feature which is not guaranteed in the solutions of NNLS or
any other methods to our knowledge. One reasonable concern that is
raised here is that nature does not know about resolution limits, so
what happens when adjacent T\textsubscript{2} components really are too
close together? Unfortunately, the answer is that we might never resolve
these two peaks in a solution where they are bound to degenerate into
one peak due to the limited SNR, as Istratov et al. \cite{istratov1999exponential}
pointed out about this particular problem: ``Many physicists have
discovered after much wasted effort that it is essential to understand
the ill-conditioned nature of the problem before attempting to compute
solutions''. On the other hand, the commonly used NNLS has incorporated
regularization techniques \cite{tihonov1963solution} to mitigate the noise
overfitting, but the choice of regularization parameters is poorly
justified in most of the MR literature. From a technical perspective,
the strength of regularization should be selected based on the local SNR
rather than being universally fixed. Unfortunately, this has never been
practiced in the NNLS analysis due to technical challenges. In addition,
the noise of a magnitude MRI image originates from Gaussian distributed
noise on both real and imaginary channels \cite{cardenas2008noise}, which is
not accounted for in the NNLS method \cite{bjarnason2010quantitative} but is
correctly modelled in the SAME-ECOS approach. Particularly in the last
few echoes, when there is no residuals of MR signal (only pure noise),
the noise profile would follow a Rayleigh distribution on the magnitude
image that always has positive values, such that NNLS could misinterpret
the Rayleigh noise as a long T\textsubscript{2} component that has not
been completely decayed yet. This phenomenon is observed in
\textbf{Figure 6}, where NNLS mean spectra (red) for all ROIs always
give a little rise on the far side of long T\textsubscript{2}s
(T\textsubscript{2} = 2000ms), but SAME-ECOS mean spectra (blue) is free
of this problem.

SAME-ECOS also takes advantage of the strong predictive power of a
fine-tuned neural network. For regression problems such as our
T\textsubscript{2}-fitting task, modern machine learning methods like
neural networks usually outperform conventional statistical methods in
terms of better prediction accuracy and faster data processing
speed \cite{hornik1989multilayer}, which is exactly what we have observed in our
simulation tests. Out of various machine learning methods, the neural
network approach is favored for use in the current SAME-ECOS workflow
because it has been successfully implemented in similar tasks by
different research groups. \cite{cohen2018mr,liu2020myelin_NN,lee2020artificial} In theory, other
machine learning methods may also achieve a similar predictive power.
Comparisons between different machine learning methods are beyond the
scope of this paper, but could be an area of future investigation.

Finally, it is worth highlighting that the simulated training dataset of
SAME-ECOS is not informed by any prior knowledge (e.g. a 2-pool model is
informed by the T\textsubscript{2} times of myelin water and IE water,
or a typical range of the MWF values obtained from previous studies).
Instead, SAME-ECOS is completely driven by a large number of random
simulations, which are only regulated by the experimental conditions
such as SNR. This approach should be valued and favored because it is
absolutely immune to (1) the potential errors of previous findings where
the prior information is acquired from; (2) the biases that prior
information may introduce into the analysis results. Simply speaking, an
informed model is more likely to perform the analysis with a bias
naturally towards the prior information. Similar to NNLS, no prior
information being needed is a desirable and specially designed feature
of SAME-ECOS.

\subsection{Disadvantages of SAME-ECOS}

A major concern of SAME-ECOS is that its fitting results are not easily
verified mathematically due to the use of a neural network. SAME-ECOS is
capable of producing reliable results only on data that are similarly
distributed as the training data. SAME-ECOS may become unpredictable and
yield uninterpretable results when applied to unfamiliar data. Although
we have generated a large training dataset by the simulations to account
for many types of variations, there is no doubt that real experimental
data have far more complexity in them. Factors such as artifacts are
detrimental to any analysis methods, but SAME-ECOS is potentially less
predictable. Unlike SAME-ECOS which empirically analyzes the data,
conventional methods process the data either analytically or
numerically, making them somewhat more mathematically explainable and
predictable when encountering these problems.

Another limitation of SAME-ECOS is related to standardization. The
variety of parameters such as the T\textsubscript{2} range, the noise
profiles, the number of neural network hidden layers etc. can be
manually selected at the user's will, which offers flexibility and
customizability, but it is difficult to propose a standard SAME-ECOS
model that works universally. The effects of changing these parameters
warrants further investigation. Although SAME-ECOS demonstrated
excellent performance in our simulation tests and in-vivo example
application, users should further validate SAME-ECOS on their own
experimental data before replacing any conventional analysis methods.

\subsection{Other quantitative MRI techniques and beyond}
Within MRI, the usage of SAME-ECOS is not just limited to multi-echo
relaxation sequences. As long as the spins are trackable by simulations,
then the SAME-ECOS methodology should also be applicable to other
quantitative MRI techniques, such as multi-component driven equilibrium
single pulse observation of T\textsubscript{1} and T\textsubscript{2}
(mcDESPOT) \cite{deoni2008gleaning} and neurite
orientation dispersion and density imaging (NODDI) \cite{zhang2012noddi},
by modifying the current
workflow accordingly. Beyond MRI, it is also possible to apply SAME-ECOS
to any quantitative sciences that involve multi-exponential decays. In
general, we believe a simulation-derived solver like SAME-ECOS is an
alternative way to produce at least comparable results to conventional
methods. It may deliver better performance, especially when analytical
and numerical solutions start to fail due to factors such as a limited
amount of data points and low SNR. Nevertheless, the SAME-ECOS
methodology seems to have the potential to be generalized for the
analysis of decay data within and beyond the MRI field.

\section{Conclusion}
We have introduced a novel method SAME-ECOS, which can decompose
multi-exponential MR relaxation data into a T\textsubscript{2} spectrum.
SAME-ECOS is highly tailorable to different experimental conditions and
various analysis models. Compared with the commonly used method NNLS,
our results have demonstrated that SAME-ECOS can yield much more
reliable T\textsubscript{2} spectra and MWF values in a dramatically
shorter processing time, by utilizing information theory and machine
learning simultaneously.

\section*{acknowledgements}
We thank the study participants and the excellent MRI technologists at UBC MRI Research Center. Funding support was provided by the Multiple Sclerosis Society of Canada, Natural Sciences and Engineering Research Council Discovery Grant.
%Acknowledgements should include contributions from anyone who does not meet the criteria for authorship (for example, to recognize contributions from people who provided technical help, collation of data, writing assistance, acquisition of funding, or a department chairperson who provided general support), as well as any funding or other support information.

%\section*{conflict of interest}

%\printendnotes

% Submissions are not required to reflect the precise reference formatting of the journal (use of italics, bold etc.), however it is important that all key elements of each reference are included.
%\bibliographystyle{plain}
\bibliography{all_reference.bib}

\begin{thebibliography}{45}
\providecommand{\natexlab}[1]{#1}
\providecommand{\url}[1]{\texttt{#1}}
\providecommand{\urlprefix}{}

\bibitem[{Whittall et~al.(1999)Whittall, Kenneth P and MacKay, Alex L and Li,
  David KB}]{whittall1999mono}
Whittall KP, MacKay AL, Li DK.
\newblock Are mono-exponential fits to a few echoes sufficient to determine T2
  relaxation for in vivo human brain?
\newblock Magnetic Resonance in Medicine: An Official Journal of the
  International Society for Magnetic Resonance in Medicine
  1999;41(6):1255--1257.

\bibitem[{Mackay et~al.(1994)Mackay, Alex and Whittall, Kenneth and Adler,
  Julian and Li, David and Paty, Donald and Graeb, Douglas}]{mackay1994vivo}
Mackay A, Whittall K, Adler J, Li D, Paty D, Graeb D.
\newblock In vivo visualization of myelin water in brain by magnetic resonance.
\newblock Magnetic resonance in medicine 1994;31(6):673--677.

\bibitem[{Istratov and Vyvenko(1999)Istratov, Andrei A and Vyvenko, Oleg
  F}]{istratov1999exponential}
Istratov AA, Vyvenko OF.
\newblock Exponential analysis in physical phenomena.
\newblock Review of Scientific Instruments 1999;70(2):1233--1257.

\bibitem[{Whittall and MacKay(1989)Whittall, Kenneth P and MacKay, Alexander
  L}]{whittall1989quantitative}
Whittall KP, MacKay AL.
\newblock Quantitative interpretation of NMR relaxation data.
\newblock Journal of Magnetic Resonance (1969) 1989;84(1):134--152.

\bibitem[{Lawson and Hanson(1995)Lawson, Charles L and Hanson, Richard
  J}]{lawson1995solving}
Lawson CL, Hanson RJ.
\newblock Solving least squares problems.
\newblock SIAM; 1995.

\bibitem[{Provencher(1982)Provencher, Stephen W}]{provencher1982contin}
Provencher SW.
\newblock CONTIN: a general purpose constrained regularization program for
  inverting noisy linear algebraic and integral equations.
\newblock Computer Physics Communications 1982;27(3):229--242.

\bibitem[{Kroeker and Henkelman(1986)Kroeker, Randall M and Henkelman, R
  Mark}]{kroeker1986analysis}
Kroeker RM, Henkelman RM.
\newblock Analysis of biological NMR relaxation data with continuous
  distributions of relaxation times.
\newblock Journal of Magnetic Resonance (1969) 1986;69(2):218--235.

\bibitem[{Du et~al.(2007)Du, Yiping P and Chu, Renxin and Hwang, Dosik and
  Brown, Mark S and Kleinschmidt-DeMasters, Bette K and Singel, Debra and
  Simon, Jack H}]{du2007fast}
Du YP, Chu R, Hwang D, Brown MS, Kleinschmidt-DeMasters BK, Singel D, et~al.
\newblock Fast multislice mapping of the myelin water fraction using
  multicompartment analysis of T decay at 3T: A preliminary postmortem study.
\newblock Magnetic Resonance in Medicine: An Official Journal of the
  International Society for Magnetic Resonance in Medicine 2007;58(5):865--870.

\bibitem[{Akhondi-Asl et~al.(2014)Akhondi-Asl, Alireza and Afacan, Onur and
  Mulkern, Robert V and Warfield, Simon K}]{akhondi2014t}
Akhondi-Asl A, Afacan O, Mulkern RV, Warfield SK.
\newblock T2-Relaxometry for Myelin Water Fraction Extraction Using Wald
  Distribution and Extended Phase Graph.
\newblock In: International Conference on Medical Image Computing and
  Computer-Assisted Intervention Springer; 2014. p. 145--152.

\bibitem[{Bertero et~al.(1982)Bertero, M and Boccacci, P and Pike, Edward
  Roy}]{bertero1982recovery}
Bertero M, Boccacci P, Pike ER.
\newblock On the recovery and resolution of exponential relaxation rates from
  experimental data: a singular-value analysis of the Laplace transform
  inversion in the presence of noise.
\newblock Proceedings of the Royal Society of London A Mathematical and
  Physical Sciences 1982;383(1784):15--29.

\bibitem[{McWhirter and Pike(1978)McWhirter, JG and Pike, E
  Rf}]{mcwhirter1978numerical}
McWhirter J, Pike ER.
\newblock On the numerical inversion of the Laplace transform and similar
  Fredholm integral equations of the first kind.
\newblock Journal of Physics A: Mathematical and General 1978;11(9):1729.

\bibitem[{Link et~al.(1991)Link, Norbert and Bauer, Siegfried and Ploss,
  Bernd}]{link1991analysis}
Link N, Bauer S, Ploss B.
\newblock Analysis of signals from superposed relaxation processes.
\newblock Journal of applied physics 1991;69(5):2759--2767.

\bibitem[{Schmidhuber(2015)Schmidhuber, J{\"u}rgen}]{schmidhuber2015deep}
Schmidhuber J.
\newblock Deep learning in neural networks: An overview.
\newblock Neural networks 2015;61:85--117.

\bibitem[{Lundervold and Lundervold(2019)Lundervold, Alexander Selvikv{\aa}g
  and Lundervold, Arvid}]{lundervold2019overview}
Lundervold AS, Lundervold A.
\newblock An overview of deep learning in medical imaging focusing on MRI.
\newblock Zeitschrift f{\"u}r Medizinische Physik 2019;29(2):102--127.

\bibitem[{Cohen et~al.(2018)Cohen, Ouri and Zhu, Bo and Rosen, Matthew
  S}]{cohen2018mr}
Cohen O, Zhu B, Rosen MS.
\newblock MR fingerprinting deep reconstruction network (DRONE).
\newblock Magnetic resonance in medicine 2018;80(3):885--894.

\bibitem[{Liu et~al.(2020)Liu, Hanwen and Xiang, Qing-San and Tam, Roger and
  Dvorak, Adam V and MacKay, Alex L and Kolind, Shannon H and Traboulsee,
  Anthony and Vavasour, Irene M and Li, David KB and Kramer, John K and
  others}]{liu2020myelin_NN}
Liu H, Xiang QS, Tam R, Dvorak AV, MacKay AL, Kolind SH, et~al.
\newblock Myelin water imaging data analysis in less than one minute.
\newblock NeuroImage 2020;210:116551.

\bibitem[{Hornik et~al.(1989)Hornik, Kurt and Stinchcombe, Maxwell and White,
  Halbert and others}]{hornik1989multilayer}
Hornik K, Stinchcombe M, White H, et~al.
\newblock Multilayer feedforward networks are universal approximators.
\newblock Neural networks 1989;2(5):359--366.

\bibitem[{Prasloski et~al.(2012)Prasloski, Thomas and Rauscher, Alexander and
  MacKay, Alex L and Hodgson, Madeleine and Vavasour, Irene M and Laule, Corree
  and M{\"a}dler, Burkhard}]{prasloski2012rapid}
Prasloski T, Rauscher A, MacKay AL, Hodgson M, Vavasour IM, Laule C, et~al.
\newblock Rapid whole cerebrum myelin water imaging using a 3D GRASE sequence.
\newblock Neuroimage 2012;63(1):533--539.

\bibitem[{Deshmane et~al.(2012)Deshmane, Anagha and Gulani, Vikas and Griswold,
  Mark A and Seiberlich, Nicole}]{deshmane2012parallel}
Deshmane A, Gulani V, Griswold MA, Seiberlich N.
\newblock Parallel MR imaging.
\newblock Journal of Magnetic Resonance Imaging 2012;36(1):55--72.

\bibitem[{Shapiro et~al.(1984)Shapiro, Finley R and Senturia, Stephen D and
  Adler, David}]{shapiro1984use}
Shapiro FR, Senturia SD, Adler D.
\newblock The use of linear predictive modeling for the analysis of transients
  from experiments on semiconductor defects.
\newblock Journal of applied physics 1984;55(10):3453--3459.

\bibitem[{Dobaczewski et~al.(1994)Dobaczewski, L and Kaczor, P and Hawkins, ID
  and Peaker, AR}]{dobaczewski1994laplace}
Dobaczewski L, Kaczor P, Hawkins I, Peaker A.
\newblock Laplace transform deep-level transient spectroscopic studies of
  defects in semiconductors.
\newblock Journal of applied physics 1994;76(1):194--198.

\bibitem[{Thomasson and Clark~Jr(1974)Thomasson, WM and Clark Jr,
  JW}]{thomasson1974analysis}
Thomasson W, Clark~Jr J.
\newblock Analysis of exponential decay curves: A three-step scheme for
  computing exponents.
\newblock Mathematical Biosciences 1974;22:179--195.

\bibitem[{Hennig(1988)Hennig, J{\"u}rgen}]{hennig1988multiecho}
Hennig J.
\newblock Multiecho imaging sequences with low refocusing flip angles.
\newblock Journal of Magnetic Resonance (1969) 1988;78(3):397--407.

\bibitem[{Prasloski et~al.(2012)Prasloski, Thomas and M{\"a}dler, Burkhard and
  Xiang, Qing-San and MacKay, Alex and Jones,
  Craig}]{prasloski2012applications}
Prasloski T, M{\"a}dler B, Xiang QS, MacKay A, Jones C.
\newblock Applications of stimulated echo correction to multicomponent T2
  analysis.
\newblock Magnetic resonance in medicine 2012;67(6):1803--1814.

\bibitem[{C{\'a}rdenas-Blanco et~al.(2008)C{\'a}rdenas-Blanco, Arturo and
  Tejos, Cristian and Irarrazaval, Pablo and Cameron, Ian}]{cardenas2008noise}
C{\'a}rdenas-Blanco A, Tejos C, Irarrazaval P, Cameron I.
\newblock Noise in magnitude magnetic resonance images.
\newblock Concepts in Magnetic Resonance Part A: An Educational Journal
  2008;32(6):409--416.

\bibitem[{Klambauer et~al.(2017)Klambauer, G{\"u}nter and Unterthiner, Thomas
  and Mayr, Andreas and Hochreiter, Sepp}]{klambauer2017self}
Klambauer G, Unterthiner T, Mayr A, Hochreiter S.
\newblock Self-normalizing neural networks.
\newblock In: Advances in neural information processing systems; 2017. p.
  971--980.

\bibitem[{Bishop(2006)Bishop, Christopher M}]{bishop2006pattern}
Bishop CM.
\newblock Pattern recognition and machine learning.
\newblock springer; 2006.

\bibitem[{Kingma and Ba(2014)Kingma, Diederik P and Ba, Jimmy}]{kingma2014adam}
Kingma DP, Ba J.
\newblock Adam: A method for stochastic optimization.
\newblock arXiv preprint arXiv:14126980 2014;.

\bibitem[{LeCun et~al.(2015)LeCun, Yann and Bengio, Yoshua and Hinton,
  Geoffrey}]{lecun2015deep}
LeCun Y, Bengio Y, Hinton G.
\newblock Deep learning.
\newblock nature 2015;521(7553):436--444.

\bibitem[{Abadi et~al.(2016)Abadi, Mart{\'\i}n and Agarwal, Ashish and Barham,
  Paul and Brevdo, Eugene and Chen, Zhifeng and Citro, Craig and Corrado, Greg
  S and Davis, Andy and Dean, Jeffrey and Devin, Matthieu and
  others}]{abadi2016tensorflow}
Abadi M, Agarwal A, Barham P, Brevdo E, Chen Z, Citro C, et~al.
\newblock Tensorflow: Large-scale machine learning on heterogeneous distributed
  systems.
\newblock arXiv preprint arXiv:160304467 2016;.

\bibitem[{Dvorak et~al.(2019)Dvorak, Adam V and Ljungberg, Emil and Vavasour,
  Irene M and Liu, Hanwen and Johnson, Poljanka and Rauscher, Alexander and
  Kramer, John LK and Tam, Roger and Li, David KB and Laule, Cornelia and
  others}]{dvorak2019rapid}
Dvorak AV, Ljungberg E, Vavasour IM, Liu H, Johnson P, Rauscher A, et~al.
\newblock Rapid myelin water imaging for the assessment of cervical spinal cord
  myelin damage.
\newblock NeuroImage: Clinical 2019;23:101896.

\bibitem[{Liu et~al.(2019)Liu, Hanwen and Rubino, Cristina and Dvorak, Adam V
  and Jarrett, Michael and Ljungberg, Emil and Vavasour, Irene M and Lee, Lisa
  Eunyoung and Kolind, Shannon H and MacMillan, Erin L and Traboulsee, Anthony
  and others}]{liu2019myelin}
Liu H, Rubino C, Dvorak AV, Jarrett M, Ljungberg E, Vavasour IM, et~al.
\newblock Myelin water atlas: a template for myelin distribution in the brain.
\newblock Journal of Neuroimaging 2019;29(6):699--706.

\bibitem[{Liu et~al.(2020)Liu, Hanwen and Ljungberg, Emil and Dvorak, Adam V
  and Lee, Lisa Eunyoung and Yik, Jackie T and MacMillan, Erin L and Barlow,
  Laura and Li, David KB and Traboulsee, Anthony and Kolind, Shannon H and
  others}]{liu2020myelin}
Liu H, Ljungberg E, Dvorak AV, Lee LE, Yik JT, MacMillan EL, et~al.
\newblock Myelin water fraction and intra/extracellular water geometric mean T2
  normative atlases for the cervical spinal cord from 3T MRI.
\newblock Journal of Neuroimaging 2020;30(1):50--57.

\bibitem[{Lee et~al.(2018)Lee, Lisa Eunyoung and Ljungberg, Emil and Shin,
  Dongmyung and Figley, Chase R and Vavasour, Irene M and Rauscher, Alexander
  and Cohen-Adad, Julien and Li, David KB and Traboulsee, Anthony L and MacKay,
  Alex L and others}]{lee2018inter}
Lee LE, Ljungberg E, Shin D, Figley CR, Vavasour IM, Rauscher A, et~al.
\newblock Inter-vendor reproducibility of myelin water imaging using a 3D
  gradient and spin echo sequence.
\newblock Frontiers in Neuroscience 2018;12:854.

\bibitem[{Ljungberg et~al.(2017)Ljungberg, Emil and Vavasour, Irene and Tam,
  Roger and Yoo, Youngjin and Rauscher, Alexander and Li, David KB and
  Traboulsee, Anthony and MacKay, Alex and Kolind,
  Shannon}]{ljungberg2017rapid}
Ljungberg E, Vavasour I, Tam R, Yoo Y, Rauscher A, Li DK, et~al.
\newblock Rapid myelin water imaging in human cervical spinal cord.
\newblock Magnetic resonance in medicine 2017;78(4):1482--1487.

\bibitem[{Han et~al.(2011)Han, Jiawei and Pei, Jian and Kamber,
  Micheline}]{han2011data}
Han J, Pei J, Kamber M.
\newblock Data mining: concepts and techniques.
\newblock Elsevier; 2011.

\bibitem[{Zhang et~al.(2001)Zhang, Yongyue and Brady, Michael and Smith,
  Stephen}]{zhang2001segmentation}
Zhang Y, Brady M, Smith S.
\newblock Segmentation of brain MR images through a hidden Markov random field
  model and the expectation-maximization algorithm.
\newblock IEEE transactions on medical imaging 2001;20(1):45--57.

\bibitem[{Laule et~al.(2008)Laule, Cornelia and Kozlowski, Piotr and Leung,
  Esther and Li, David KB and MacKay, Alex L and Moore, GR
  Wayne}]{laule2008myelin}
Laule C, Kozlowski P, Leung E, Li DK, MacKay AL, Moore GW.
\newblock Myelin water imaging of multiple sclerosis at 7 T: correlations with
  histopathology.
\newblock Neuroimage 2008;40(4):1575--1580.

\bibitem[{Laule et~al.(2006)Laule, C and Leung, E and Li, D KB and Traboulsee,
  AL and Paty, DW and MacKay, AL and Moore, G RW}]{laule2006myelin}
Laule C, Leung E, Li DK, Traboulsee A, Paty D, MacKay A, et~al.
\newblock Myelin water imaging in multiple sclerosis: quantitative correlations
  with histopathology.
\newblock Multiple Sclerosis Journal 2006;12(6):747--753.

\bibitem[{Celik et~al.(2013)Celik, Hasan and Bouhrara, Mustapha and Reiter,
  David A and Fishbein, Kenneth W and Spencer, Richard
  G}]{celik2013stabilization}
Celik H, Bouhrara M, Reiter DA, Fishbein KW, Spencer RG.
\newblock Stabilization of the inverse Laplace transform of multiexponential
  decay through introduction of a second dimension.
\newblock Journal of Magnetic Resonance 2013;236:134--139.

\bibitem[{Tihonov(1963)Tihonov, Andrei Nikolajevits}]{tihonov1963solution}
Tihonov AN.
\newblock Solution of incorrectly formulated problems and the regularization
  method.
\newblock Soviet Math 1963;4:1035--1038.

\bibitem[{Bjarnason et~al.(2010)Bjarnason, Thorarin A and McCreary, Cheryl R
  and Dunn, Jeff F and Mitchell, J Ross}]{bjarnason2010quantitative}
Bjarnason TA, McCreary CR, Dunn JF, Mitchell JR.
\newblock Quantitative T2 analysis: the effects of noise, regularization, and
  multivoxel approaches.
\newblock Magnetic Resonance in Medicine: An Official Journal of the
  International Society for Magnetic Resonance in Medicine 2010;63(1):212--217.

\bibitem[{Lee et~al.(2020)Lee, Jieun and Lee, Doohee and Choi, Joon Yul and
  Shin, Dongmyung and Shin, Hyeong-Geol and Lee, Jongho}]{lee2020artificial}
Lee J, Lee D, Choi JY, Shin D, Shin HG, Lee J.
\newblock Artificial neural network for myelin water imaging.
\newblock Magnetic Resonance in Medicine 2020;83(5):1875--1883.

\bibitem[{Deoni et~al.(2008)Deoni, Sean CL and Rutt, Brian K and Arun, Tarunya
  and Pierpaoli, Carlo and Jones, Derek K}]{deoni2008gleaning}
Deoni SC, Rutt BK, Arun T, Pierpaoli C, Jones DK.
\newblock Gleaning multicomponent T1 and T2 information from steady-state
  imaging data.
\newblock Magnetic Resonance in Medicine: An Official Journal of the
  International Society for Magnetic Resonance in Medicine
  2008;60(6):1372--1387.

\bibitem[{Zhang et~al.(2012)Zhang, Hui and Schneider, Torben and
  Wheeler-Kingshott, Claudia A and Alexander, Daniel C}]{zhang2012noddi}
Zhang H, Schneider T, Wheeler-Kingshott CA, Alexander DC.
\newblock NODDI: practical in vivo neurite orientation dispersion and density
  imaging of the human brain.
\newblock Neuroimage 2012;61(4):1000--1016.

\end{thebibliography}

\end{document}